%% file: icse.tex
\documentclass[sigconf]{acmart}

\acmConference[ICSE 2022]{The 44th International Conference on Software Engineering}{May 21–29, 2022}{Pittsburgh, PA, USA}

\input{packages}
\input{macros}

\setcopyright{acmcopyright}
\copyrightyear{2022}
\acmYear{2022}
\acmDOI{10.1145/1122445.1122456}

\begin{document}

\title{Fuzzing Class Specifications}


\author{Facundo Molina}
\affiliation{%
  \institution{University of Rio Cuarto and CONICET}
  \country{Argentina}}
\email{fmolina@dc.exa.unrc.edu.ar}

\author{Marcelo d'Amorim}
\affiliation{%
  \institution{Federal University of Pernambuco}
  \country{Brazil}}
\email{damorim@cin.ufpe.br}

\author{Nazareno Aguirre}
\affiliation{%
  \institution{University of Rio Cuarto and CONICET}
  \country{Argentina}}
\email{naguirre@dc.exa.unrc.edu.ar}

\renewcommand{\shortauthors}{F.~Molina, M. d'Amorim and N. Aguirre}

\input{abstract}

\keywords{Oracle problem, specification inference, grammar-based fuzzing}

\maketitle

\input{introduction}

\input{background}

\input{illustrative-examples}
\input{technique}
\input{evaluation}
\input{related-work}
\input{conclusion}

\section*{Acknowledgements}
This work is partially supported by INES (\url{www.ines.org.br}); CNPq grant 465614/2014-0; CAPES grant 88887.136410/2017-00; FACEPE grants APQ-0399-1.03/17 and PRONEX APQ/0388-1.03/14; and ANPCyT grants PICT 2017-2622 and PICT 2019-2050. Facundo Molina’s work is also supported by Microsoft Research, through a Latin America PhD Award.

Any opinions, findings, and conclusions or recommendations expressed in this publication are those of the authors, and do not necessarily reflect the views of the sponsoring entities.
\bibliographystyle{ACM-Reference-Format}
\bibliography{references}

\end{document}

%% file: packages.tex
\usepackage[group-separator={,}, group-minimum-digits={4}]{siunitx}
\usepackage{wrapfig}
\usepackage[font=small,labelfont=bf]{caption}
\usepackage{pbox}
\usepackage{syntax}
\usepackage{amsmath}
\usepackage{tablefootnote}
\usepackage{gensymb}
\usepackage{epsfig}
\usepackage{enumerate}
\usepackage{xspace}
\usepackage{epsf,picinpar}
\usepackage{varioref}
\usepackage{colortbl,multirow,hhline}
\usepackage{listings}
\usepackage{colortbl,multirow,hhline}
\usepackage{algorithmic}
\usepackage{algorithm}
\usepackage[normalem]{ulem}
\usepackage{pifont}
\usepackage{url}
\usepackage{arydshln}
\makeatletter
\g@addto@macro{\UrlBreaks}{\UrlOrds}
\makeatother
\usepackage{balance}
\usepackage{graphicx, subfigure}
\usepackage{longtable}
\usepackage{lscape}
\usepackage{multirow}
\usepackage{listings}
\usepackage{framed}
\usepackage{morefloats}
\usepackage[T1]{fontenc}
\usepackage{pdfpages}
\usepackage{fancybox}
\usepackage{adjustbox}
\usepackage{flushend}
\usepackage{booktabs}
\usepackage{enumitem}
\usepackage{tabularx}
\usepackage{diagbox}
\usepackage{soul} 
\usepackage{booktabs, multirow} 
\usepackage{soul}
\usepackage{changepage,threeparttable} 
\usepackage{placeins} 
\usepackage[most]{tcolorbox}
\usepackage{graphicx}
\usepackage{tikz}
\usepackage{pgf-pie}   
\usepackage{longtable}
\usepackage{dcolumn}
\usepackage{xltabular}
\usepackage{marginnote}
\usetikzlibrary{calc,matrix,chains,positioning,decorations.pathreplacing,arrows}

\usepackage{listings}
\usepackage{xcolor}
\usepackage{color}

\definecolor{darkblue}{rgb}{0,0,0.5}
\definecolor{LightGray}{rgb}{0.75,0.75,0.75}
\definecolor{VeryLightGray}{rgb}{0.90,0.90,0.90}
\definecolor{ForestGreen}{rgb}{0.13,0.54,0.13}
\definecolor{lightgreen}{rgb}{0.8,1,0.8}
\definecolor{lightyellow}{rgb}{1,1,0.8}
\definecolor{lightred}{rgb}{1,0.8,0.8}

\lstdefinelanguage{Python}{
    morecomment=[f][\lstbg{lightgreen}]+,
    morecomment=[f][\lstbg{lightred}]-,
    morecomment=[f][\lstbg{lightyellow}]\\,
}
\newcommand{\lstbg}[3][0pt]{{\fboxsep#1\colorbox{#2}{\strut 3}}}

%% file: macros.tex
\newcolumntype{P}[1]{>{\raggedright\arraybackslash}p{#1}}

\newcolumntype{H}{>{\setbox0=\hbox\bgroup}c<{\egroup}@{}}
\newcolumntype{Z}{>{\setbox0=\hbox\bgroup}c<{\egroup}@{\hspace*{-\tabcolsep}}}

\definecolor{gray50}{gray}{.5}
\definecolor{gray40}{gray}{.6}
\definecolor{gray30}{gray}{.7}
\definecolor{gray20}{gray}{.8}
\definecolor{gray10}{gray}{.9}
\definecolor{gray05}{gray}{.95}

\definecolor{celadon}{rgb}{0.67, 0.88, 0.69}
\definecolor{lightcoral}{rgb}{0.94, 0.5, 0.5}

\definecolor{black}{rgb}{0.2, 0.2, 0.2}
\definecolor{darkgreen}{rgb}{0, 0.5, 0}

\definecolor{lightgrey}{rgb}{0.9, 0.9, 0.9}

\newcommand{\ie}{i.e.\xspace}

\newcommand{\PreserveBackslash}[1]{\let\temp=\\#1\let\\=\temp}
\newcolumntype{C}[1]{>{\PreserveBackslash\centering}p{#1}}
\newcolumntype{R}[1]{>{\PreserveBackslash\raggedleft}p{#1}}
\newcolumntype{L}[1]{>{\PreserveBackslash\raggedright}p{#1}}

\makeatletter
\newcommand\footnoteref[1]{\protected@xdef\@thefnmark{\ref{#1}}\@footnotemark}

\makeatother

\newcommand{\Comment}[1]{} 

\definecolor{dkgreen}{rgb}{0,0.6,0}
\definecolor{gray}{rgb}{0.5,0.5,0.5}
\definecolor{mauve}{rgb}{0.58,0,0.82}

\definecolor{mygray}{gray}{1}

\newcommand{\tname}{\textsc{SpecFuzzer}}
\newcommand{\CodeIn}[1]{{\footnotesize \texttt{#1}}}
\newcommand{\codetexttt}[1]{{\small \texttt{#1}}}

\newcommand{\repoURL}{\textbf{\url{https://sites.google.com/view/specfuzzer}}}

\newcommand{\programp}{$\mathcal{P}$}
\newcommand{\programpoint}{$\rho$}

%% file: abstract.tex
\begin{abstract}
Expressing class specifications via executable constraints is important for various software engineering tasks such as test generation, bug finding and automated debugging, but developers rarely write them. Techniques that infer specifications from code exist to fill this gap, but they are designed to support specific kinds of assertions and are difficult to adapt to support different assertion languages, e.g., to add support for quantification, or additional comparison operators, such as membership or containment.   

\sloppy
To address the above issue, we present \tname{}, a novel technique that combines grammar-based fuzzing, dynamic invariant detection, and mutation analysis, to automatically produce class specifications. \tname{} uses: \emph{(i)} a fuzzer as a generator of candidate assertions derived from a grammar that is automatically obtained from the class definition; \emph{(ii)} a dynamic invariant detector --Daikon-- to filter out assertions invalidated by a test suite; and \emph{(iii)} a mutation-based mechanism to cluster and rank assertions, so that similar constraints are grouped and then the stronger prioritized. Grammar-based fuzzing enables \tname{} to be straightforwardly adapted to support different specification languages, by manipulating the fuzzing grammar, e.g., to include additional operators. 

We evaluate our technique on a benchmark of 43 Java methods employed in the evaluation of the state-of-the-art techniques GAssert and EvoSpex. Our results show that \tname{} can easily support a more expressive assertion language, over which is more effective than GAssert and EvoSpex in inferring specifications, according to standard performance metrics. 
\end{abstract}

%% file: introduction.tex
\section{Introduction}


Software specifications are abstract descriptions of the software's intended behavior. They serve two main purposes: to explicitly state the user needs and to check implementation conformance~\cite{Ghezzi+2002}. In Object-Oriented (OO) design, where software is organized as a set of classes, a class specification describes the intended behavior of the class methods and the constraints on state.
While the specification of a class is typically described \emph{informally}, through natural language documentation of its API, the specification becomes significantly more useful when expressed \emph{formally}, through constraints known as \emph{contracts}\footnote{In the context of this paper, we will interchangeably use the terms \emph{contract} and \emph{specification}. A contract is typically composed of different assertions for various program points, such as method preconditions and postconditions.} \cite{Meyer1997,Schiller-TypeContracts}.
Contracts enabled techniques of various kinds, including test generation \cite{DamorimETAL-ASE2006,Tillmann+2008}, automated debugging \cite{Demsky+2006, Perkins+2009, Logozzo+2012}, bug finding \cite{Pacheco+2007,Cok+2005}, and verification \cite{Cok+2005,Furia+2017}.

Techniques for inferring class specifications exist~\cite{Blassi+2018,DBLP:journals/scp/ErnstPGMPTX07,gassert2020,Molina+2021}, but their expressiveness is limited. Daikon \cite{DBLP:journals/scp/ErnstPGMPTX07}, the baseline that other techniques use, supports a restricted set of templates, from which assertions are generated. It is then limited to simple assertions (e.g., no direct support for quantification), or requires the developer to manually extend the assertion language. GAssert \cite{gassert2020} and EvoSpex \cite{Molina+2021}, two recently proposed techniques for contract inference, try to address this limitation of Daikon by supporting more expressive assertion languages, but their extensions focus on specific kinds of constraints: GAssert focuses on logical/arithmetic constraints (no quantified expressions) and EvoSpex focuses on object navigation constraints (only very simple logical and arithmetic operators are supported). Moreover, as both techniques are based on evolutionary search, they are difficult to extend or adapt to support further expressions, as the evolutionary algorithms are targeted for the specific languages supported by the corresponding tools. 

To overcome the limitations of existing approaches, we propose \tname{}, a technique for generating likely specifications by \emph{fuzzing} potential specifications associated with a given class. \tname{} uses grammar-based fuzzing to automatically generate constraints that can be used as candidate specifications by an invariant detection tool (in our case, we use Daikon). Fuzzing~\cite{fuzzingbook2019:index}, traditionally used to efficiently produce structured random data for testing, has two key advantages in this context: (1)~it eliminates the need of developers to manually define candidate assertions and (2)~it enables developers to straightforwardly adapt the language of assertions by manipulating the fuzzing grammar. 

Fuzzing can quickly produce very large sets of assertions to be fed to a dynamic detection tool. However, as the assertions are generated randomly, and dynamic invariant detection only filters out assertions that can be invalidated by a given test suite, a substantial number of candidate specification expressions may be reported by a fuzzer. To address that problem and better assist developers in driving their attention to the likely most relevant specification, \tname{} uses a prioritization mechanism based on clustering and mutation testing~\cite{DBLP:journals/ac/PapadakisK00TH19,DBLP:books/daglib/0020331}. After generating thousands of candidate specifications with fuzzing, \tname{} uses the output of a custom mutation analysis to cluster candidate specifications. More precisely, it partitions the set of specifications according to the mutants they kill, and within each partition, the assertion that is falsified the most number of times when running the test suite on the mutants, is picked as the representative. Notice that, even though all the assertions in the same partition kill the same mutants, some may be falsified more than others, as a same mutant may be killed by multiple tests. The rationale for the mutation based partition is that assertions that kill different mutants are non-equivalent (or, alternatively, that assertions that kill the same mutants are ``similar''); the rationale for ranking assertions according to the number of failures is that assertions that are falsified a greater number of times are ``stronger''.

We compared \tname{} with GAssert~\cite{gassert2020}~and EvoSpex~\cite{Molina+2021}, which are the state of the art tool-supported techniques in specification inference today. To evaluate \tname, we used the same benchmarks from the evaluation of GAssert and EvoSpex, carefully studied the subjects, and manually produced corresponding ``ground truth'' assertions capturing the intended behavior of the subjects. We then used this ground truth to accurately assess precision and recall of \tname{}, GAssert, and EvoSpex. It is worth noting that (1) prior work used indirect metrics to compute precision and recall (as opposed to the direct usage of ground truth) and that (2) prior work used a subset of subjects we consider (our benchmark is the combination of the GAssert and EvoSpex benchmarks). Our results show that \tname{} increases the expressiveness over GAssert and EvoSpex, being able to express $\sim$45\% more assertions in the ground truth than these related tools. \tname{} was also able to detect 75\% of all assertions in the ground truth, showing a better overall performance compared to previous techniques. The results we obtained provide initial, yet strong evidence that \tname\ is effective.


In summary, this paper makes the following contributions:
\begin{itemize}
    \item \tname{}, a novel technique for assertion inference combining grammar-based fuzzing and dynamic invariant detection.
    \item An efficient mechanism for grouping similar assertions and for ranking assertions based on their strength. 
    \item A thorough evaluation of our technique, in comparison with the related tools GAssert and EvoSpex, where performance metrics are computed in relation to manually written ground truth assertions. 
\end{itemize}

The evaluation artifacts of \tname\ are publicly available for download: \repoURL{}.

%% file: background.tex
\section{Background}

This section presents background material that is important for the rest of the paper.


\subsection{Specification Inference}
\label{spec-inference}


Specification inference is the problem of generating a formal description of the software behavior from existing software artifacts, e.g., documentation, source code, etc. Specification inference is closely related to the \emph{oracle problem}~\cite{Barr+2015}, which is the problem of deciding whether or not a program execution is consistent with the desired behavior of the program. Specification inference provides a means to create \emph{oracles}~\cite{Barr+2015}. For regression testing purposes, it sometimes suffices to produce specifications of expected properties as assertions for the context of a given test case~\cite{DBLP:conf/issta/FraserZ10}. However, more general assertions that capture properties \emph{at given locations within the program (not the test) for any input} have other applications, including testing. This is the problem we study in this paper, defined as follows.

\begin{definition}
\label{task:problem}
Given a target program \programp{}, and a program point of interest \programpoint{} in \programp{}, infer a specification $\phi$ that captures the states at \programpoint{}, \ie{}, for every state $s$ of \programp{}, $\phi$ holds in $s$ if and only if there exists an execution $t$ of \programp{} such that $s$ is the state of $t$ at program point \programpoint{}. 
\end{definition}

\subsection{Grammar-based Fuzzing}
\label{grammar-based-fuzzing}


\sloppy
Fuzzing is a very active topic both in research~\cite{relatedness-fuzzing-tools} and practice~\cite{clusterfuzz,oss-fuzz,afl}. Fuzzing is a technique to automatically produce large sets of (often structured) data, for testing a target program. The generation process typically involves randomness and the rationale is that testing on (large sets of) quasi-valid data can reveal subtle bugs, such as wrongly handled inputs and corner cases. A well-known use case of fuzzing is detection of security vulnerabilities, such as buffer overflows~\cite{miller-fuzzing,manes-tse-survey2019}. 

Different fuzzing strategies exist~\cite{manes-tse-survey2019}. Grammar-based fuzzing uses an input grammar to produce syntactically-valid inputs by traversing the production rules of the grammar. In its simplest form, the input generation process can be implemented as an incremental expansion of a string starting from the initial grammar symbol, and replacing non-terminal symbols by the application of a randomly-chosen production rule of the corresponding non-terminals, until the string consists of terminals only; a bound on the number of non-terminals enables this process to handle recursion, which would otherwise lead to infinite loops. As an example, consider a scenario where the program to test takes as input a propositional logic (PL) formula, characterized by the grammar from Figure~\ref{prop-grammar}. To generate testing data, the PL grammar can be fed to a grammar-based fuzzer (e.g., Grammarinator~\cite{10.1145/3278186.3278193}) to efficiently obtain a very large set of well-formed test data (PL formulas, in this case). To generate inputs, the fuzzer explores paths induced by the grammar production rules. For instance, the input \texttt{neg(p and q)} can be obtained through the following derivation: $\mathit{start\leadsto{}formula\leadsto{}}$neg $\mathit{formula} \leadsto{}$ neg $(\mathit{formula}$ and $\mathit{formula})\leadsto{}$ neg (p and $\mathit{formula})\leadsto{}$ neg (p and q). It is worth noticing that a great advantage of fuzzing in this case is that the input language can be easily adapted by modifying the grammar. For instance, our fuzzer would be able to generate formulas with disjunctions if we add a corresponding production rule to the non-terminal $\mathit{formula}$.

\setlength{\grammarparsep}{0.1cm} 
\setlength{\grammarindent}{1cm} 
\begin{figure}[t!]
\begin{center}
\begin{grammar}
<start> ::= <formula>

<formula> ::= <atomic> | neg <formula> | (<formula> and <formula>)

<atomic> ::== true | false | p | q | r | ...
\end{grammar}
\end{center}
\caption{\label{prop-grammar}Propositional Logic grammar.}
\vspace{-3ex}
\end{figure}

The above example is relevant because the technique we propose in this paper uses grammar-based fuzzing as a lightweight approach to produce assertions (such as the PL formulas above) as candidate specifications for program points. The simplicity with which the grammar can be adapted or extended will be one of the advantages of the approach, compared with related techniques. 

\subsection{Assertion Language}
\label{assertion-language}

An assertion is a logical expression associated with a program point expressing an expected property at that location. The use of assertions has wide-spread applications in software design \cite{Meyer1997}, software testing \cite{DBLP:books/daglib/0020331}, and verification~\cite{DBLP:journals/annals/Hoare03,DBLP:journals/sigsoft/ClarkeR06}.

Our assertion language is similar in expressive power to JML, including first-order quantification (\codetexttt{\textbackslash{}forall}, \codetexttt{\textbackslash{}exists}), arithmetic and logical operators, and reachability expressions (the reach operator \codetexttt{\textbackslash{}reach(x, f1, \dots, fk)} denotes the smallest set of objects reachable from \codetexttt{x}, through fields \codetexttt{f1}, \dots, \codetexttt{fk}). Additionally, postcondition assertions might use the \codetexttt{\textbackslash{}old(expr)} notation, to refer to the value of expression \codetexttt{expr} at the precondition. For simplicity, we drop the backslashes, shorten the quantifier names, and replace the semicolon notation in JML quantification by the implication (in the case of universal quantification) or conjunction (in the case of existential quantification). As an example, the following expression

{\footnotesize
\begin{verbatim}
all SList l: reach(this, next).has(l) ==> l.elem == old(l.elem)
\end{verbatim}
}

\noindent
states that the integer field \CodeIn{elem} of the list nodes reachable from \texttt{this} (\texttt{has} is the JML operator for membership) remains unchanged. This expression corresponds to the following JML expression:

{\footnotesize
\begin{verbatim}
\forall SList l; \reach(this, next).has(l); l.elem == \old(l.elem)
\end{verbatim}
}

Our assertion language is motivated by the expressive power of the languages in related work, and in contract languages \cite{DBLP:journals/scp/ErnstPGMPTX07,gassert2020,Molina+2021,Meyer1997,DBLP:conf/fmco/ChalinKLP05,DBLP:conf/sas/Fahndrich10}. This is a \emph{general} language, that includes the usual relational, arithmetic and logical operators, but no domain specific functions (e.g., trigonometric functions, that would be relevant only for some analysis subjects, are not considered). The assertion language enables one to refer to class/object fields, but not to the results of method calls. That would require us to declare methods as ``pure'' to use them in assertions, which is beyond what our current implementation supports.


%% file: illustrative-examples.tex
\begin{figure}[t!]
  \begin{lstlisting}[xleftmargin=0cm,xrightmargin=3.6cm]
/* Returns the minimum of two integers */
public static int min(int x, int y) {
  if (x <= y) return x;
  else return y;
}
\end{lstlisting}
\vspace{-8ex}
\caption{\label{get-min}Method to get the minimum of two values.}
\end{figure}

\begin{figure}[t!]
  \begin{lstlisting}[xleftmargin=0cm,xrightmargin=3.6cm]
public class SList {
  private int elem; 
  private SList next;
  private static final int SENTINEL = Integer.MAX_VALUE;
  
  /* Constructors */
  public SList() { this(SENTINEL, null); }
  private SList(int elem, SList next) {
    this.elem = elem; 
    this.next = next; 
  }
  
  /* Method to insert an element in the list */
  void insert(int data) {
    if (data > elem) {
      next.insert(data);
    } else {
      next = new SList(elem, next); 
      elem = data; 
    }
  } }
\end{lstlisting}
\vspace{-8ex}
\caption{\label{sorted-list}Class \codetexttt{SList} implements an ordered list of integers.}
\vspace{-3ex}
\end{figure}

\section{Illustrative Examples}
\label{illustrative}

This section illustrates \tname{} on two simple examples with the purpose of (1) highlighting limitations of state-of-the-art specification inference techniques and (2) illustrate \tname.

\subsubsection*{Examples} 

Figure~\ref{get-min} shows \codetexttt{min}, a Java method to compute the minimum of two integers, whereas Figure~\ref{sorted-list} shows \codetexttt{SList}, a Java class implementing an ordered list of integers. The \codetexttt{min} method is straightforward. Class \codetexttt{SList} is slightly more elaborate. It has two instance fields, \codetexttt{elem} and \codetexttt{next}, that represent the value of a linked list node and the reference to the next node, respectively. It also has a class field (\codetexttt{SENTINEL}) that stores a special value --the maximum Java integer value-- as a mark for the end of the list. The sentinel should be placed at the end of the list and should not be repeated. The default constructor creates a node marking the end of the list. The \codetexttt{insert} method takes the integer \codetexttt{data} as parameter and inserts it in its right-sorted position in the linked list. As it is not possible for any integer value to be greater than the sentinel, the search is guaranteed to insert the element before the sentinel.


\subsubsection*{Relevant Properties}

The intended behavior of method \codetexttt{min} is that it computes the minimum between \codetexttt{x} and \codetexttt{y}. A specification of the postcondition of \codetexttt{min} in our assertion language is as follows:

{\footnotesize
\begin{verbatim}
(result == x || result == y) && (result <= x) && (result <= y)
\end{verbatim}
}

The postcondition of method \codetexttt{SortedList.insert} involves various properties: the list is acyclic and sorted increasingly, the sentinel is in the list (at the end), and the \codetexttt{data} element is inserted. This postcondition can be specified as follows:

{\footnotesize
\begin{verbatim}
all SList l: reach(this, next).has(l) ==> !reach(l.next, next).has(l)
all SList l: reach(this, next).has(l) ==> l.elem <= l.next.elem
exists SList l: reach(this, next).has(l) && l.elem == SENTINEL
exists SList l: reach(this, next).has(l) && l.elem == data
\end{verbatim}
}

We may consider these assertions to be the \emph{ground truth} postcondition specifications of the corresponding methods, and what we would ideally expect specification inference tools to produce. 

\begin{table*}[ht!]
\caption{Daikon, GAssert, EvoSpex and \tname{} on the running examples.}
\label{techniques-on-examples}
\centering
\renewcommand{\arraystretch}{1.2} 
\scriptsize
\setlength\dashlinedash{1pt}
\begin{tabular}{ll|l|l|l}
\hline 
& \multicolumn{1}{c}{\small \textbf{Daikon}} & \multicolumn{1}{c}{\small \textbf{GAssert}} & \multicolumn{1}{c}{\small \textbf{EvoSpex}} & \multicolumn{1}{c}{\small \textbf{\tname{}}} \\
\hline
\multicolumn{5}{c}{\cellcolor{gray05} \textbf{min(int x,int y) - postcondition}} \\
\hline
1 & \CodeIn{r <= x} & \CodeIn{(x > r \&\& y == r) ||} & \CodeIn{r <= x} & \CodeIn{x >= r} \\
& & \CodeIn{(r <= y \&\& r == x)} & & \\
2 & \CodeIn{r <= y} & & \CodeIn{r <= y} & \CodeIn{x < r ==> r <= 1} \\
3 & & & & \CodeIn{x >= y ==> y == r} \\
4 & & & & \CodeIn{x >= y || y != r} \\
5 & & & & \CodeIn{x <= y || y <= r} \\
6 & & & & \CodeIn{x <= y || y >= r} \\
7 & & & & \CodeIn{x >= y <==> y == r} \\
8 & & & & \CodeIn{x == y ==> y <= r} \\
9 & & & & \CodeIn{x <= y <==> x == r} \\
\hline 
\multicolumn{5}{c}{\cellcolor{gray05} \textbf{SList.insert(int data) - postcondition}} \\
\hline 
1 & \CodeIn{e <= next.e} & \CodeIn{e-(data-e) <= old(e)} & \CodeIn{exists SList l: reach(this, next).has(l)}  & \CodeIn{exists SList l: reach(this, next).has(l)} \\
& & & \CodeIn{   \&\& l.e == data} & \CodeIn{   \&\& l.e == SENTINEL} \\
2 & \CodeIn{e <= next.next.e} & & \CodeIn{old(e) <= next.e} & \CodeIn{all SList l: reach(this, next).has(l)} \\
& & & & \CodeIn{   ==> l.e <= l.next.e} \\
3 & \CodeIn{next.e <=} & & & \CodeIn{exists SList l: reach(this, next).has(l)} \\
& \CodeIn{   next.next.e} & & & \CodeIn{   \&\& l.e == data} \\
4 & \CodeIn{next != null} & & & \CodeIn{next != null} \\
5 & \CodeIn{e <= old(e)} & & & \CodeIn{e != old(e) + 1} \\
6 & \CodeIn{e <= old(next.e)} & & & \CodeIn{next.e >= old(e)} \\
7 & \CodeIn{e <=} & & & \CodeIn{data >= next.e || next.e = old(e)} \\
& \CodeIn{   old(next.next.e)} & & & \\
8 & \CodeIn{e <= data} & & & \CodeIn{e = data xor data > old(e)} \\
9 & \CodeIn{next.e >= old(e)} & & & \CodeIn{e > data ==> data = next.e} \\
10 & \CodeIn{next.e <=} & & & \CodeIn{exists SList l: reach(this.next, next).has(l)} \\
& \CodeIn{   old(next.e)} & & & \CodeIn{   \&\& l.e != 1} \\
11 & \CodeIn{next.e <=} & & & \CodeIn{exists SList l: reach(this, next).has(l)} \\
& \CodeIn{   old(next.next.e)} & & & \CodeIn{   \&\& l.e > this.e} \\
12 & \CodeIn{next.next.e >=} & & & \CodeIn{exists SList l: reach(this, next).has(l)} \\
& \CodeIn{   old(e)} & & & \CodeIn{   \&\& l.e <= l.next.e} \\
13 & \CodeIn{next.next.e >=} & & & \CodeIn{all SList l: reach(this.next, next).has(l)} \\
& \CodeIn{   old(next.e)} & & & \CodeIn{   ==> l.e >= this.next.e} \\
14 & \CodeIn{next.next.e <=} & & & \CodeIn{e <= old(e) ==> old(e) < old(next.next.e)} \\
& \CodeIn{   old(next.next.e)} & & & \\
15 & & & & \CodeIn{e != next.next.e + old(next.next.e)} \\
16 & & & & \CodeIn{data >= next.next.e ||} \\
& & & & \CodeIn{next.next.e = old(next.e)} \\
\hline
\end{tabular}
\end{table*}

\subsection{Techniques for Specification Inference}

\subsubsection*{Daikon}

Daikon~\cite{DBLP:journals/scp/ErnstPGMPTX07} is dynamic technique that infers specifications by monitoring test executions. Considering Definition~\ref{task:problem}, besides the program \programp{}, Daikon requires a test suite ${\mathcal T}$ for \programp{} to infer specifications. Daikon uses ${\mathcal T}$ to exercise \programp{}; it monitors program states at various program points of \programp{}; it considers a set of assertions obtained by instantiating assertion patterns, and those that are not invalidated by any test at a given program point are reported to the user as \emph{likely invariants} at the program point. 

\subsubsection*{GAssert and EvoSpex}

GAssert~\cite{gassert2020} and EvoSpex~\cite{Molina+2021} are recently proposed specification inference techniques. As Daikon, these tools execute a test suite of the program under analysis and observe executions to \emph{infer} specifications that are consistent with the observations. While Daikon requires the test suite to be provided, GAssert and EvoSpex use their own test generation mechanisms (third-party test generation tools in the case of GAssert, a custom test generation approach in the case of EvoSpex). Although both techniques are based on evolutionary search, they have key differences. GAssert implements a co-evolutionary algorithm that explores the space of possible assertions (the co-evolution deals with false-positives and false-negatives via two cooperating evolutionary processes) and uses the OASIs \cite{DBLP:conf/issta/JahangirovaCHT16} oracle assessment tool to iteratively improve the assertions. EvoSpex implements a classical genetic algorithm to explore the search space, and uses a state mutation technique to generate postcondition states in which the assertions being sought for should fail. GAssert's evolutionary operations focus on logical and arithmetic assertions whereas EvoSpex's focuses on object navigational properties. For these tools, changing the assertion languages implies redefining the corresponding evolutionary operators and other parameters of the evolutionary algorithms, which is non-trivial. 

Table~\ref{techniques-on-examples} shows how Daikon, GAssert, and EvoSpex perform on the examples (for brevity, we have used \codetexttt{e} instead of \codetexttt{elem}, and removed \codetexttt{this} from non-quantified expressions in the \codetexttt{insert} example). GAssert performs perfectly on the \codetexttt{min} example, but poorly on \codetexttt{SList} (it does not capture most of the ground truth); EvoSpex infers one complex assertion for \codetexttt{SList.insert} (that the element is inserted) and misses the remaining three in the corresponding ground truth; it also infers part of the ground truth for \codetexttt{min}. Daikon infers the same as EvoSpex in the case of \codetexttt{min}, and in the case of \codetexttt{SList.insert}, it only infers specific sortedness instances between the first few elements of the list, but it fails to generalize this relationship for the whole structure. It fully misses the remaining assertions in the ground truth.

\subsection{\tname\ }

\tname{} uses a combination of static analysis, grammar-based fuzzing, and mutation analysis to infer specifications. \tname\ proceeds as follows. First, it uses a lightweight static analysis to produce a grammar for the specification language, which is tuned to the software under analysis. Then, it uses a grammar-based fuzzer to generate candidate specifications from that grammar. A dynamic detector then determines which of those specifications are consistent with the behavior exhibited by a provided test suite. 
Finally, \tname\ eliminates irrelevant and equivalent specifications using a mechanism based on mutation analysis and clustering. A salient feature of \tname\ is that developers can adjust the set of specifications produced by tuning the grammar as opposed to making changes in the tool.


Table~\ref{techniques-on-examples} shows the assertions that \tname{} infers as postconditions for methods \codetexttt{min} and \codetexttt{SList.insert}. Recall that \tname\ uses fuzzing and reports a higher number of assertions compared to the other techniques. We configured the fuzzer to produce 2000 candidate assertions per subject and found that of those 51 and 437 were confirmed as likely invariants by the dynamic detector for \codetexttt{min} and \codetexttt{insert}, respectively. The mutation-based partition strategy is what allows \tname{} to considerably reduce the reported assertions to 9 and 16, respectively. 

In the case of \codetexttt{min}, the 9 inferred assertions are valid and their conjunction is equivalent to the corresponding ground truth. For \codetexttt{SList.insert}, the first 3 assertions already cover 3 out of 4 assertions in the ground truth (the only missing one is list acyclicity). The other inferred assertions are either valid but less relevant (4-13), or invalid (14-16). The invalid ones are specifications that were true in the provided test suite, but there exist some unseen scenarios in which they are falsified. Notice that this also affects the other techniques, even though GAssert and EvoSpex include costly mechanisms to reduce invalid assertions (the assertion inferred by GAssert for \codetexttt{SList.insert}, in particular, is an invalid property).

%% file: technique.tex
\begin{figure*}[t!]
\begin{center}
\includegraphics[width=1.8\columnwidth]{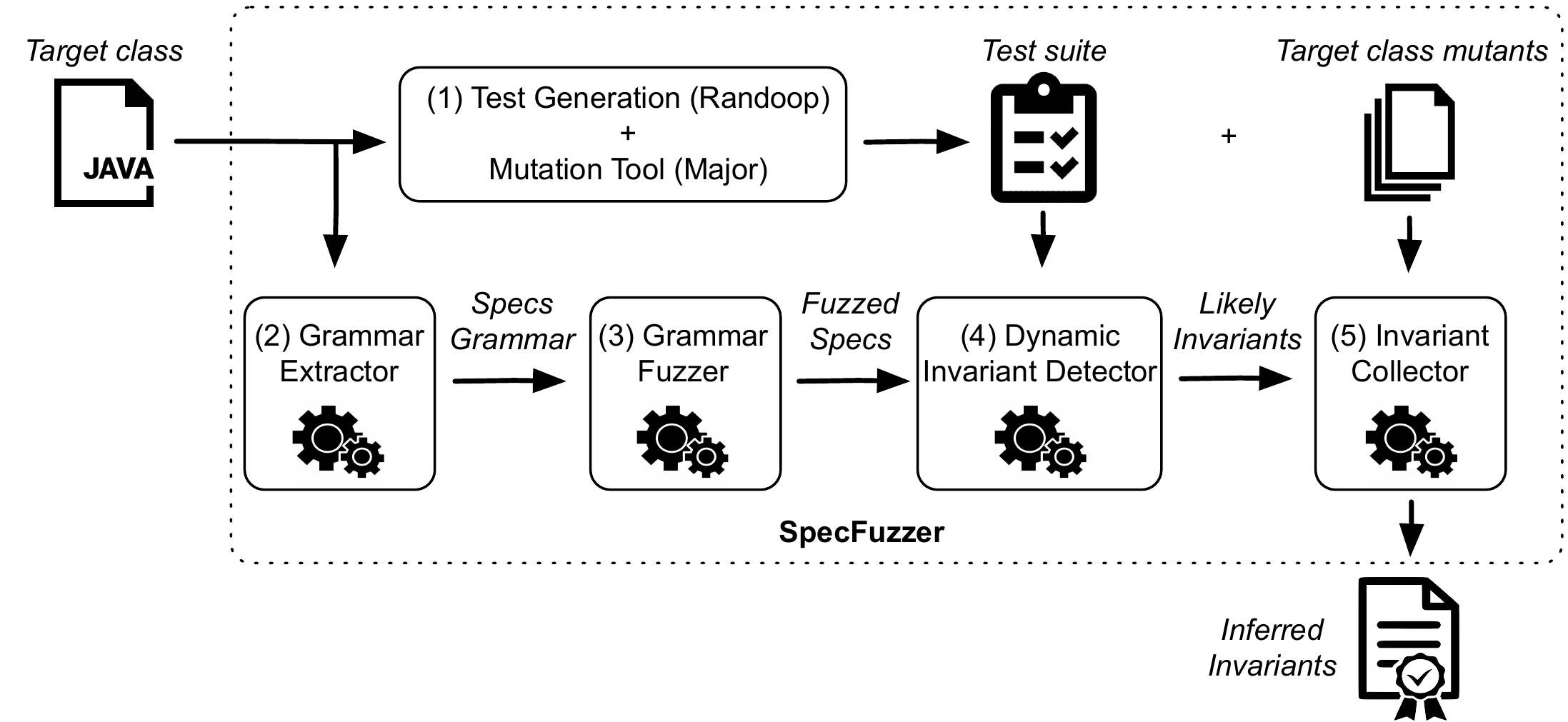}
\end{center}
\vspace{-2ex}
\caption{\label{workflow}The \tname\ workflow.}
\label{overview-approach}
\end{figure*}

\section{Approach}

This section presents \tname{}, a technique for specification inference that uses a combination of static analysis, grammar-based fuzzing, and mutation analysis. 

Figure~\ref{overview-approach} shows the workflow of the technique. \tname{} takes as input a Java\footnote{Although our approach is general and language independent, some parts of our current prototype, such as the grammar extraction and the evaluation of candidate (fuzzed) specifications, are currently implemented for Java. Supporting other languages that Daikon can handle, like C, would require the implementation of such parts.} class $C$, and produces assertions that seek to characterize properties of different execution points in $C$, such as method preconditions and postconditions. Following the Daikon terminology, we will generally refer to assertions that hold on specific program points as \emph{invariants}. The technique is organized as a pipeline of five components: (1) a \emph{Tests and Mutants Generation} component that produces tests and mutants for other components of the pipeline, (2) a \emph{Grammar Extractor} that analyzes $C$ to generate a specification grammar for that class, (3) a \emph{Grammar Fuzzer}, which produces candidate assertions by exploring the production rules from the extracted grammar, (4) a \emph{Dynamic Invariant Detector}, responsible for inferring likely invariants from the fuzzed assertions via observations made with the executions of an input test suite, and (5) an \emph{Invariant Selector} component, which partitions the likely invariants produced by the previous component to discard useless (weak) assertions, groups together similar assertions, and reports a reduced set of assertions, prioritizing the stronger ones. The following sections discuss these components in greater detail.


\subsection{Tests and Mutants Generation}
\label{test-gen-and-mutation-tool}

The first step of our process to infer specifications for a class $C$ consists of \emph{(i)} generating a test suite $T$ exercising the methods of the target class $C$, and \emph{(ii)} producing a set $M_{1}, \dots, M_{n}$ of mutants of $C$, representing synthetic faults in the class. As Figure~\ref{workflow} shows, these artifacts are used at different stages of the technique. We used Randoop\footnote{\url{https://randoop.github.io/randoop/}} for test generation and Major\footnote{\url{https://mutation-testing.org/}} for mutant generation. Although we used these tools in our current implementation, the user may replace them with
other tools or even provide her own test suite and mutated versions of the target class $C$.



\subsection{Grammar Extractor}
\label{grammar-extraction}

\setlength{\grammarparsep}{0.1cm} 
\setlength{\grammarindent}{1cm} 

\begin{figure}[t]
\begin{center}
\begin{grammar}
<FuzzedSpec> ::= <QuantifiedExpr> | <BooleanExpr>

<QuantifiedExpr> ::= <Quantifier> <Typed\_Var> `:' <BooleanExpr> 

<Quantifier> ::= `all' | `exists'

<BooleanExpr> ::= <NumCmpExpr> | <LogicCmpExpr> | <MembershipExpr> | `!'  <BooleanExpr>

<NumCmpExpr> ::= <NumExpr> <NumCmpOp> <NumExpr> \alt
	<NumExpr> <NumCmpOp> <NumExpr> <NumBinOp> <NumExpr>

<NumExpr> ::= <NumVar> | <NumConst>





<LogicCmpExpr> ::= <BooleanExpr> <LogicOp> <NumCmpExpr> \alt
    `(' <BoolVar> <LogicOp> <BoolVar> `)' <LogicOp> <NumCmpExpr> \alt
    `(' <NumCmpExpr> `)' <LogicOp> `(' <NumCmpExpr> `)' 


<MembershipExpr> ::= <type\_SetExpr>.has(<type\_Var>) 

<NumCmpOp> ::= `==' | `!=' | `>' | `<' | `<=' | `>='

<NumBinOp> ::= `+' | `-' | `*' | `/' | `\%'

<LogicOp> :== `||' | `xor' | `==>' | `<==>'
\end{grammar}
\end{center}
\caption{Fragment of the base grammar $B$.}
\label{base-grammar}
\end{figure}


The \textit{Grammar Extractor} takes as input a class $C$ and creates a grammar $G_{C}$ expressing the language of candidate assertions for $C$.
Those assertions denote method preconditions, postconditions, and class invariants. The extractor instantiates our base grammar, referred to as $B$, with information that is specific to $C$, e.g., attribute types, legally typed navigational expressions involving the attributes, etc. 


Figure~\ref{base-grammar} shows a fragment of the base grammar $B$, capturing the fixed parts of the specification language, \ie{}, the parts that are common to any input class of interest. For this paper, the grammar $B$ supports numerical comparisons, logical expressions, membership expressions, and quantified expressions. Numerical comparisons and logical expressions are the simplest constructs of the language. They relate numerical expressions and boolean expressions by using traditional numerical operators and logical connectives---$\langle \mathit{NumCmpOp} \rangle$ and $\langle \mathit{LogicOp} \rangle$, respectively. Membership expressions allow one to express whether or not a typed element belongs to a set (collection) of the corresponding type. The grammar fragment uses the \texttt{has} notation from JML, and shows a production rule for typed variables. Although it is not explicitly shown in the grammar fragment, the \texttt{reach} operator is a way of building a typed set expression. A concrete example of a membership expression from a formula shown in Section~\ref{illustrative} is the following:

{\small
\begin{verbatim}
    reach(this, next).has(l)
\end{verbatim}
}

\noindent
expressing that list \codetexttt{l} belongs to the set of objects reachable from 
\codetexttt{this} by navigations (zero or more) through \codetexttt{next}. Finally, the grammar allows for existential and universal quantification. Again, an example of a quantified expression from Section~\ref{illustrative} is the following:

{\small
\begin{verbatim}
exists SList l:
     reach(this, next).has(l) && l.elem == SENTINEL 
\end{verbatim}
}

\noindent
whose intuitive reading is that there exists a list object reachable from \codetexttt{this} with the field \CodeIn{elem} holding the \codetexttt{SENTINEL}.

To obtain the grammar $G_{C}$, the grammar extractor takes $B$ and adds or deletes symbols and production rules, based on the structure of $C$. The process basically depends on $C$'s direct and indirect fields (fields declared in $C$ itself or in a class reachable from $C$). Intuitively, from every field/navigation, a terminal symbol of the corresponding type is defined (e.g., \codetexttt{this.next} will be a terminal of type \codetexttt{SList}). 

Set expressions deserve a more detailed description. Firstly, if a field \codetexttt{f} is of a \codetexttt{Collection} type, then \codetexttt{f} will be a terminal of (typed) \codetexttt{SetExpr}. For instance, if \codetexttt{SList} were an implementation of \codetexttt{Collection}, then \texttt{this} and \codetexttt{this.next} would be terminals of type \codetexttt{SList\_SetExpr}. Secondly, the \codetexttt{reach} operator is also involved in building set expressions. For expression \codetexttt{e} and recursive field \codetexttt{f} (a field is recursive if it is defined in a class \codetexttt{C} and has type \codetexttt{C}) of class \codetexttt{C}, a production rule allows expression \codetexttt{reach(e, f)} to have type \codetexttt{C\_SetExpr}. Thus, expression \codetexttt{reach(this, next)} has type \codetexttt{SList\_SetExpr}.

\subsection{Grammar Fuzzer}

\begin{figure}[t]
    \begin{center}
        \includegraphics[scale=0.45]{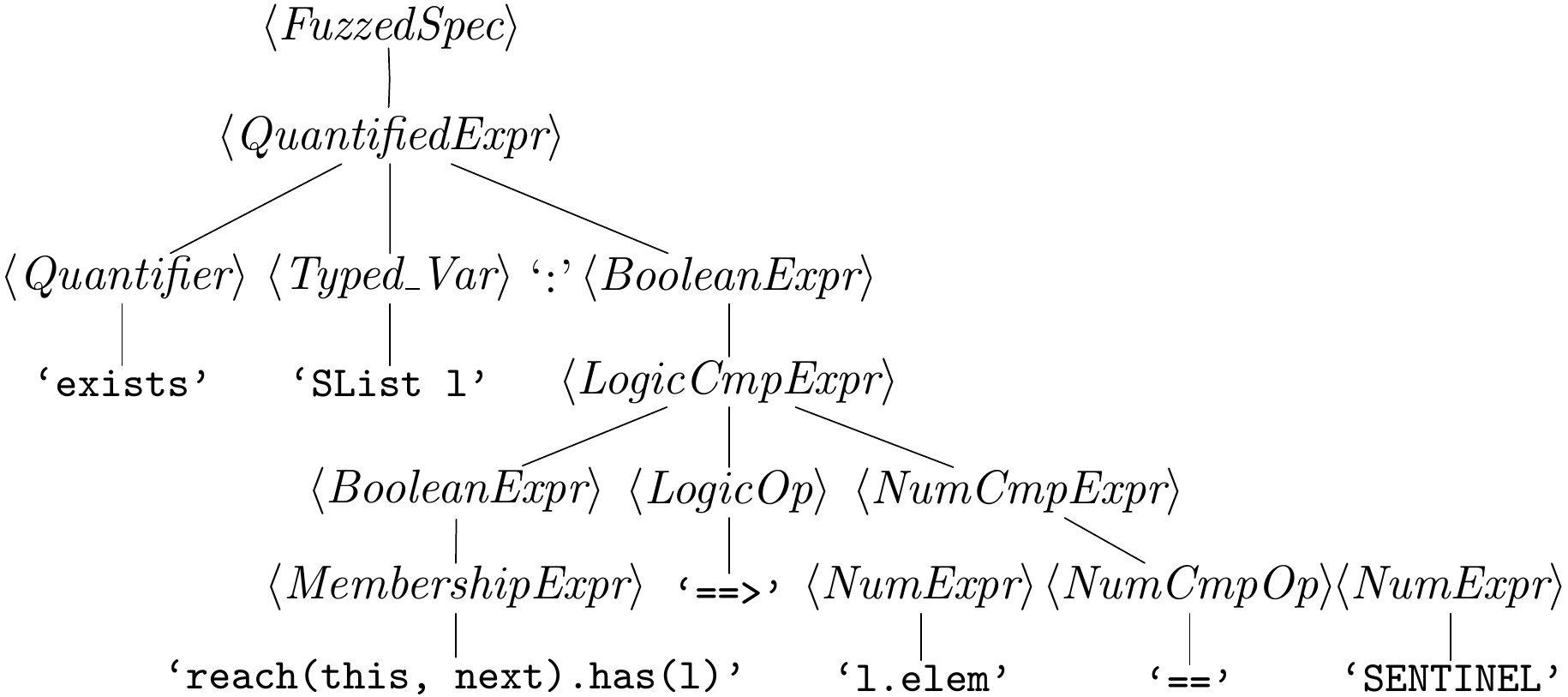}
    \end{center}
    \caption{A derivation tree produced by our Grammar Fuzzer for the expression \codetexttt{exists SList l: reach(this, next).has(l) \&\& l.elem == SENTINEL}.}
    \label{derivation-tree}
\end{figure}

The goal of the \emph{Grammar Fuzzer} component is to produce candidate assertions. It uses a standard generative grammar-based fuzzer to achieve this goal~\cite{10.1145/3278186.3278193,fuzzingbook2019:index}. This component produces derivations of $G_{c}$---\ie{}, strings in $\mathcal{L}(G_{c})$---to obtain assertions for $C$. It begins with the start symbol $\langle$\emph{FuzzedSpec}$\rangle$ and keeps expanding non-terminal symbols until no more non-terminals are present. Each non-terminal symbol is expanded based on a non-deterministic choice and, to avoid expansions leading to infinite derivation paths, a limit of 5 on the number of non-terminals is defined. Furthermore, to avoid getting stuck in a situation where the number of symbols cannot be reduced any further, the total number of expansion steps is also limited to 100. The rationale for this choice is that complex class assertions can be created by combining small assertions, rather than longer ones. Figure~\ref{derivation-tree} shows the derivation tree of the property used in our illustrative example: \codetexttt{exists SList l: reach(this, next).has(l) \&\& l.elem == SENTINEL}.

By using this derivation mechanism, our \emph{Grammar Fuzzer} produces candidate predicates very efficiently. In all of our experiments we generated up to 2,000 different candidates every time we executed \tname{}. Furthermore, as the grammar $G_{C}$ has been specifically extracted for a class $C$, all the specifications generated by the fuzzer are guaranteed to express properties over $C$. We implemented our fuzzer in Java, reproducing a general grammar-based fuzzer written in Python~\cite{fuzzingbook2019:index}.\footnote{https://www.fuzzingbook.org/html/Grammars.html\#A-Simple-Grammar-Fuzzer}



\subsection{Dynamic Invariant Detector}
\label{inference-step}



The goal of the \emph{Dynamic Invariant Detector} is to evaluate the plausibility of the candidate assertions produced by the fuzzer. As Figure~\ref{overview-approach} shows, the dynamic invariant detector takes as input a test suite, produced by the test generator, and a set of assertions, produced by the fuzzer. This component instruments the program with the assertions generated by the fuzzer and runs the tests to verify which assertions hold across all executions. The resulting assertions are reported as \emph{likely invariants}.

The dynamic invariant detector is built on top of Daikon,\footnote{http://plse.cs.washington.edu/daikon/} a state-of-the-art tool for likely invariant detection~\cite{DBLP:journals/scp/ErnstPGMPTX07}. We used Daikon as follows. We configured Daikon to include the assertions we provided---\ie{}, the expressions produced by the fuzzer---in the initial pool of candidate assertions it uses. For that, we used a mechanism provided by Daikon to incorporate new constraints.\footnote{http://plse.cs.washington.edu/daikon/download/doc/developer.html\#New-invariants} Furthermore, we included, along with the new constraints, a component that allows the tool to interpret and evaluate the assertions at run-time. 

\subsection{Invariant Selector}
\label{filtering-step}


The goal of the \emph{Invariant Selector} is to partition the assertions that were deemed valid by the dynamic invariant detector, grouping together similar assertions, and taking a single representative from each partition. At the same time, this component discards assertions that, although were confirmed by the invariant detector, are considered weak and thus less relevant. This component takes as input the set of likely invariants obtained from the previous step, and the set of mutants of the input class \codetexttt{C} (obtained from a mutation tool). This component reports a subset of the likely assertions passes as input, ranking the invariants by the number of failures in corresponding code assertions. The Invariant Selector reduces the number of reported assertions. To sum up, this component discards an assertion because of one of two reasons:
\begin{enumerate}
    \item the assertion is considered \emph{weak}, and not to capture relevant properties of the target class;
    \item the assertion is (semantically) \emph{similar} to another produced assertion. 
\end{enumerate}

In the following, we describe how we approximate the detection of weak and equivalent specifications via mutation analysis.

\subsubsection{Detecting weak specifications with Mutation Analysis}
\label{sec:detect-weak-specification}


Recall that the fuzzer reports thousands of constraints and the dynamic invariant detector (in our case, Daikon) can only discard specifications invalidated by the tests. Several constraints can still ``survive'' the filtering process described on Section~\ref{inference-step}. Even with better test suites some assertions would still "survive" that process. For example a \emph{tautology}, such as \codetexttt{x >= y || x <= y}, would \emph{not} be invalidated by Daikon as it is a valid proposition, but it is unlikely to be useful. Being syntactically driven, the fuzzer can produce valid assertions that do not provide any interesting information. The assumption is that it also generates interesting ones. \tname{} uses \emph{mutation analysis to discard uninteresting assertions}. The idea is the following: if an assertion --one that is not falsified by the test suite of \codetexttt{C}-- cannot be falsified by any mutant of \codetexttt{C}, then it is a property that not only holds on \texttt{C}, but also on all synthetic buggy versions of \texttt{C}. We will then consider it a \emph{weak} assertion, and discard it as being \emph{irrelevant}. This approach of using mutation analysis to induce more effective (stronger) oracles has been used in prior work, notably by \citet{DBLP:conf/issta/FraserZ10}, as well as in recent work on oracle improvement \cite{gassert2020} and specification inference \cite{Molina+2021}. 


\subsubsection{Clustering similar specifications with Mutation Analysis}
\label{sec:detect-similar-specification}
It is possible that \tname\ produces syntactically different assertions that are semantically equivalent (or similar with respect to a distance metric). \tname\ tries to identify and \emph{remove} such assertions. As an example, consider the following assertions that are produced by \tname\  on our \codetexttt{SList} example:

{\small
\begin{verbatim}
all SList l: reach(this, next).has(l) ==> 
                                    l.elem <= l.next.elem
all SList l: !(reach(this, next).has(l) && 
                                    l.elem > l.next.elem)
\end{verbatim}
}


\begin{figure}[t!]
    \centering
    \resizebox{0.8\linewidth}{!}{
        \begin{tikzpicture}   
                \pie[cloud,text=legend,scale font, text=legend,]{24/\Huge{Irrelevant}, 72.4/\Huge{Equivalent}, 3.6/\Huge{Reported}} 
        \end{tikzpicture} 
    }
    \caption{\label{piechart}Breakdown of reasons for discarding specifications on \codetexttt{SList.insert}. Only 3.6\% of the specifications that ``survive'' the invariant detection stage are reported.}
    \vspace{-3ex}
\end{figure}
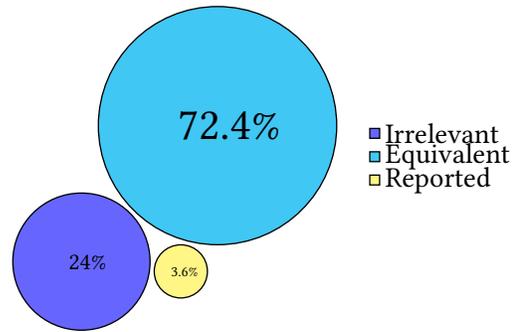

\noindent
Both these assertions express the sortedness property on lists. Their equivalence of those specs follows from De Morgan's laws~\cite{Novikov1964-NOVEOM}, and algebraic properties of integers, and the equivalence of boolean connectives. To identify equivalent assertions and assertions that are similar with respect to their ability to capture synthetic faults, \tname\ again uses mutation analysis. Two assertions will be considered similar if they kill the same set of mutants, \ie{}, if the are falsified on the same set of program faults. For example, the two assertions above kill the same set of 2 mutants, together with 74 other specifications. More precisely, \tname\ partitions the set of likely assertions according to the mutants they kill. Finally, from each partition, \tname\ selects a representative assertion. To do so, we proceed with the following heuristic: the assertions in each partition are ranked by the number of times they fail when running the test suite on the mutants (while they all kill the same mutants, some assertions may fail a greater number of times, i.e., for most tests in the test suite). The rationale is that assertions that fail the most represent stronger properties, and thus they may subsume other assertions in the partition. Considering the \codetexttt{SList.insert} example, this mechanism enabled \tname\ to reduce the number of reported specifications from 437 specifications to 16. Figure~\ref{piechart} shows the breakdown of assertions classified as irrelevant (Section~\ref{sec:detect-weak-specification}), equivalent (Section~\ref{sec:detect-similar-specification}), and reported.

\subsubsection*{Implementation} 

\tname\ is publicly available for download in the replication package.\footnote{\repoURL{}}

%% file: evaluation.tex
\section{Evaluation}

To evaluate \tname{}, we performed a series of experiments focused on the following research questions:

\begin{itemize}
    
    \item[\textbf{RQ1}] \emph{Is grammar-based fuzzing effective at generating relevant assertions?}

    \item[\textbf{RQ2}] \emph{Is the mutation-based selector successful for removing redundant/irrelevant assertions?}

    \item[\textbf{RQ3}] \emph{How does \tname{} compare with alternative techniques?}
    
\end{itemize}
RQ1 analyzes the effectiveness of using grammar-based fuzzing as a technique to generate candidate assertions, with respect to a ground truth. RQ2 evaluates the suitability of the mutation-based assertion selection component of \tname{}, at discarding weak assertions, and prioritizing the most relevant. Finally, RQ3 compares the effectiveness of \tname{} with the state-of-the-art techniques GAssert~\cite{gassert2020} and EvoSpex~\cite{Molina+2021}.

\subsection{Evaluation Subjects}

The performance evaluation in previous work used indirect metrics to compute precision and recall \cite{gassert2020,Molina+2021}. In this paper, we compute these measures directly, which requires having a ground truth for comparison, for each evaluation subject. We took all the 34 Java methods that were part of the evaluation of GAssert~\cite{gassert2020}, and all 23 methods in the contract reproducibility evaluation of EvoSpex~\cite{Molina+2021}, obtaining a data set composed of 57 subject methods. We studied each method, and manually produced corresponding ``ground truth'' assertions capturing the intended behavior of the corresponding method.  These assertions were cross-checked by authors and can be found in our replication package. Notice that, since previous work focuses only on inferring postconditions, our evaluation also focuses in these program points, although \tname{} can infer assertions for various program points. We obtained a total of 80 ground truth assertions, for the 57 methods. We then carefully examined each ground truth assertion, to determine if they can be expressed in the assertion language of at least one of the evaluated tools. Since 15 of the 80 assertions were not supported by any of these tools, we discarded them\footnote{Discarded assertions include complex trigonometric properties, vector cross product constraints, assertions involving characters and strings, and conversions between characters and hexadecimal encodings.}. Methods that, after this removal, ended up without ground truth postconditions, were also removed. This results in 65 postcondition assertions, for 43 Java methods.

Each assertion within the 65 in the ground truth could be expressed in the assertion language of at least one tool. GAssert's language can express 28 of the 65; EvoSpex's language can express 29 of the 65; and \tname{} can express 41 of the 65. Although our grammar supports all 65 assertions, when implementing support for the grammar into Daikon's assertion template instantiation, some expressions are ignored by Daikon's infrastructure (e.g., expressions which templates instantiation require objects of different classes). There is no fundamental reason why these issues can't be resolved, but they demand modifications in Daikon.

\subsection{Experimental Setup}

\tname{} requires a test suite for the class under analysis, and a set of mutants for this class. The test suite was generated using Randoop \cite{Pacheco+2007}, with a maximum of 500 test sequences to produce. Mutants were generated using Major \cite{Just+2011}, with all supported mutation operators enabled.  The fuzzer was run to produce up to 2,000 different candidate assertions (syntactic duplicates were removed), for each subject class.

Regarding GAssert and EvoSpex, we followed the same methodology described in their papers~\cite{gassert2020,Molina+2021}, using exactly the same configuration parameters for the evolutionary processes of each technique. Moreover, to account for the randomness of each approach, for each of the 43 Java methods we inferred postconditions with each tool a total of 10 times. All the results reported in this section correspond to the averages of the executions. 

We set a timeout of 90 minutes, for each execution of each tool. All tools were run on an Intel Core i7 3.2Ghz, with 16Gb of RAM, running GNU/Linux (Ubuntu 18.04). The detailed description of how to reproduce the experiments in this paper is available in the replication package site.

\subsection{Grammar-based Fuzzing Effectiveness}

The effectiveness of grammar-based fuzzing in producing relevant assertions is measured against assertions in the ground truth. The experiment for RQ1 consisted in running \tname{} on each subject, and analyzing the percentage of those assertions that the tool was able to infer. Recall that the invariant selector uses a (mutation-based) heuristic to discard assertions. As such, it may wrongly discard relevant assertions. For that reason, to answer RQ1, we ran \tname{} with the invariant selector disabled.

We manually inspected the output of \tname{}. More precisely, we manually analyzed the assertions that \tname{} reports to verify if they were present in the ground truth (and the contrary). In some cases, it was difficult to determine if the assertion was equivalent to a certain assertion in the ground truth. When the obtained expressions allowed for it, we used an SMT solver via Microsoft IntelliTest to check expression equivalence. In short, we produced C\# programs to check whether the set of inferred assertions was equivalent to assertions in the ground truth.

\begin{table}[t!]
\caption{\label{recall-fuzzing}Inferring assertions with grammar-based fuzzing.}
\vspace{-2ex}
\begin{center}
\begin{footnotesize}
\renewcommand{\arraystretch}{1.2} 
\setlength{\tabcolsep}{0.4em}
\scalebox{1.0}{
\begin{tabular}{crrr}
\toprule
\textbf{Ground-truth} & \textbf{\#Reported} & \textbf{\#Detected} & \textbf{\%Detected} \\
\midrule
65 & 20,277 & 40 & \textbf{61.5} \\
41 & 15,555 & 40 & \textbf{97.5} \\
\bottomrule
\end{tabular}}
\end{footnotesize}
\end{center}
\vspace{-1ex}
\end{table}

Table~\ref{recall-fuzzing} summarizes the results of the experiments for RQ1, with respect to the overall ground truth (65 assertions) and the subset of the ground truth that is actually supported by \tname{} (as we explained before, 41 out of the 65 are currently supported by our implementation). We report the number of reported assertions, the number of ground truth assertions detected by the tool, and the percentage of ground truth assertions that were detected. If we consider the language supported by our implementation, our tool correctly detects 97.5\% of the assertions in the ground truth; if we consider the language  supported by at least one of the specification inference tools, \tname{} correctly detects 61.5\% of the ground truth assertions. These results confirm that grammar-based fuzzing is effective in generating relevant assertions (as shown later on, even when considering the 65 assertions in the ground truth, the performance of fuzzing is comparable with state-of-the-art tools). 

\subsection{Performance of Invariant Selection}

The invariant selector component of \tname{} implements mutation-based heuristics to reduce the number of reported assertions, discarding ``weak'' assertions (assertions that survive all mutants), and selecting representatives among the assertions that kill the same mutants. RQ2 evaluates the performance of this stage. The experiment in this case compares the assertions obtained after invariant detection, with the assertions that are preserved after running the invariant selection. The comparison measures assertion reduction, and the percentage of the ground truth that is covered prior and after assertion selection. 

Table~\ref{reducedproperties} shows these results for all subjects, grouped by class name. In each case we report assertions in the ground truth, reported assertions before and after invariant selection, and percentage of the ground truth that is covered, again, before and after invariant selection. Finally, we indicate the reduction rate obtained by invariant selection (number of assertions after selection, with respect to the number of assertions before selection). 

\setlength\dashlinedash{1pt}

\newlength\tbspace
\setlength\tbspace{0.2cm}
\newcolumntype{R}{r<{\hspace{\tbspace}}}

\begin{table}[t!]
\caption{\label{reducedproperties}Performance of the Invariant Selector reducing assertions.}
\vspace{-2ex}
\begin{center}
\setlength{\tabcolsep}{0.5em}
\footnotesize
\begin{tabular}{lr|rr|rr|r}
\toprule
\multirow{2}{*}{\textbf{Subject}} &
\multirow{2}{*}{\textbf{\#GT}} &
\multicolumn{2}{c}{\textbf{\#Reported}} & 
\multicolumn{2}{c}{\textbf{Detected(\%)}} &
\multirow{2}{*}{\textbf{Red.~($\%$)}} \\
& & \textbf{Before} & \textbf{After} & \textbf{Before} & \textbf{After} & \\
\midrule
oasis.SimpleMethods & 4 & 115 & 31 & 75 & 75 & 73 \\
daikon.StackAr & 8 & 2067 & 70 & 87.5 & 62.5 & 96.6 \\
daikon.QueueAr & 8 & 4699 & 152 & 50 & 50 & 96.7 \\
math.ArithmeticUtils & 1 & 4 & 2 & 100 & 100 & 50 \\
math.FastMath & 2 & 60 & 31 & 100 & 100 & 48.3 \\
math.MathUtils & 1 & 19 & 4 & 0 & 0 & 78.9 \\
lang.BooleanUtils & 5 & 49 & 12 & 100 & 100 & 75.5 \\
guava.IntMath & 1 & 314 & 46 & 0 & 0 & 85.3 \\
tsuite.Angle & 2 & 3 & 0 & 50 & 0 & 100 \\
tsuite.MathUtil & 3 & 22 & 13 & 33.3 & 33. 3& 40.9 \\
tsuite.Envelope & 1 & 1094 & 27 & 0 & 0 & 97.5 \\
\hdashline
eiffel.Composite & 4 & 8696 & 42 & 75 & 0 & 99.5 \\
eiffel.DLLN & 3 & 137 & 29 & 100 & 100 & 78.8 \\
eiffel.Map & 6 & 140 & 22 & 16.6 & 16.6 & 84.2 \\
eiffel.RingBuffer & 5 & 1947 & 269 & 20 & 20 & 86.1 \\
cozy.Polyupdate & 3 & 382 & 119 & 66.6 & 66.6 & 68.8 \\
cozy.Structure & 2 & 153 & 21 & 100 & 100 & 86.2 \\
cozy.ListComp02 & 2 & 62 & 5 & 0 & 0 & 91.9 \\
cozy.MinFinder & 1 & 17 & 2 & 100 & 100 & 88.2 \\
cozy.MaxBag & 3 & 297 & 78 & 100 & 100 & 73.7 \\
\noalign{\vskip 0.5mm}
\hline
\noalign{\vskip 0.5mm}
TOTAL & 65 & 20277 & 975 & 61.5 & 52.3 & 95.1 \\
& 41 & 15555 & 618 & 97.5 & 82.9 & 96 \\
\bottomrule
\end{tabular}
\end{center}
\vspace{-2ex}
\end{table}

The invariant selection results show that the mutation-based heuristics in \tname{} effectively reduces the number of reported assertions, with a relatively small loss in property detection (with respect to the ground truth). More precisely, the reported assertions are reduced by $\sim$95\%, and 6 out of the 40 correctly fuzzed assertions are discarded (covering $\sim$52\% of the ground truth of 65 assertions, $\sim$83\% of the 41 ground truth assertions that the tool supports). 

To understand the reasons why we miss 6 ground truth assertions during invariant selection, we analyzed how these assertions are classified by the detector. In all cases, the assertions are deemed as \emph{irrelevant}, i.e., they are not killed by any mutant. While the problem may be a weak test suite, it becomes clear, when observing the assertions, that there is no mutation operator able to kill these assertions (the assertions are shown in Table~\ref{missed-properties}). The problem is not specific to Major (the mutation tool that we used); other tools such as PIT do not have mutants able to kill these assertions either. Let us provide two concrete examples. Assertion \CodeIn{abs(res) <= 1} for \codetexttt{Angle.getTurn} corresponds to a method whose \codetexttt{result} is either 0, -1 or 1; no mutant makes this method return a value other than these. In the \CodeIn{Composite.addChild} subject, assertion \CodeIn{c.value == old(c.value)} would be violated if a mutant changed the value of \CodeIn{c}, a parameter of the method; a mutation operator achieving this effect would have to add a new sentence.

These observations suggest that we may improve the effectiveness of our heuristics by extending Major with support for additional mutation operators, specific to our purposes. 

\subsection{Comparison of GAssert, EvoSpex and \tname{}}
\label{techniques-comparison-section}

RQ3 compares \tname{} with the state-of-the-art tools GAssert and EvoSpex. The comparison is based on standard performance metrics: precision, recall and f1-score. These metrics are computed with respect to the ground truth that we produced for the evaluation subjects. Tools were run to infer assertions as described earlier in this section, and the results are shown in Table~\ref{techniques-comparison}, grouped by subject class. Columns \#M and \#GT show the number of methods in the subject and the number of assertions in the ground truth, respectively. For each technique we show the number of inferred assertions, the precision and recall with respect to the ground truth, and the f1-score. We summarize the performance metrics for the overall ground truth of 65 assertions, as well as in the context of assertions that are supported by each particular tool (recall that GAssert supports in its language 28 of the 65, EvoSpex 29 of the 65, and \tname{} 41 of the 65). That is, rows Total-SpecFuzzer, Total-GAssert and Total-EvoSpex show the performance of the techniques on the portion of the ground truth that \tname{}, GAssert and EvoSpex support, respectively.  

\begin{table}[t!]
\caption{\label{missed-properties}Valid assertions discarded by the Invariant Selector.}
\vspace{-2ex}
\centering
\renewcommand{\arraystretch}{1.2} 
\setlength\dashlinedash{1pt}
\scalebox{1.2}{
\scriptsize
\begin{tabular}{ll}
\toprule
\textbf{Subject} & \textbf{Assertion}  \\
\midrule
StackAr.pop & \CodeIn{theArray[old(top)] == null} \\
StackAr.topAndPop & \CodeIn{theArray[old(top)] == null} \\
Angle.getTurn & \CodeIn{abs(res) <= 1} \\
Composite.addChild & \CodeIn{c.value == old(c.value))} \\
& \CodeIn{children == old(children)} \\
& \CodeIn{ancestors == old(ancestors)} \\
\bottomrule
\end{tabular}}
\vspace{-2ex}
\end{table}

\begin{table*}[]
\footnotesize
\caption{\label{techniques-comparison}Precision, Recall and F1-Score of GAssert, EvoSpex and \tname{} on the data set.}
\vspace{-2ex}
\centering
\setlength{\tabcolsep}{0.2em}
\renewcommand{\arraystretch}{1.2} 
\begin{tabular}{lrR|rrR|rrR|rrR|rrr}
\toprule
\multirow{2}{*}{\textbf{Subject}} &
\multirow{2}{*}{\textbf{\#M}} &
\multirow{2}{*}{\textbf{\#GT}} &
\multicolumn{3}{c}{\textbf{\#Inferred}} & 
\multicolumn{3}{c}{\textbf{Precision(\%)}} &
\multicolumn{3}{c}{\textbf{Recall(\%)}} & 
\multicolumn{3}{c}{\textbf{F1-Score}}\\
\cmidrule(r{\tbspace}){4-6} \cmidrule(r{\tbspace}){7-9} \cmidrule(r{\tbspace}){10-12} \cmidrule(r{\tbspace}){13-15}
& & & \textbf{GAssert} & \textbf{EvoSpex} & \textbf{SpecFuzzer} & 
\textbf{GAssert} & \textbf{EvoSpex} & \textbf{SpecFuzzer} & 
\textbf{GAssert} & \textbf{EvoSpex} & \textbf{SpecFuzzer} &
\textbf{GAssert} & \textbf{EvoSpex} & \textbf{SpecFuzzer} \\
\midrule
oasis.SimpleMethods &
4 &
4 & 
7 & 4 & 31 & 
100 & 75 & 100 & 
50 & 25 & 75 & 
0.66 & 0.37 & 0.85 
\\
daikon.StackAr &
5 &
8 & 
5 & 6 & 70 & 
100 & 83.3 & 87.1 & 
37.5 & 37.5 & 62.5 & 
0.54 & 0.51 & 0.72 
\\
daikon.QueueAr &
5 &
8 & 
8 & 12 & 152 & 
100 & 91.6 & 61.1 & 
37.5 & 25 & 50 & 
0.54 & 0.39 & 0.54 
\\
math.ArithmeticUtils &
1 &
1 & 
1 & 0 & 2 & 
100 & 100 & 50 & 
100 & 0 & 100 & 
1 & 0 & 0.66 
\\
math.FastMath &
1 &
2 & 
3 & 1 & 31 & 
100 & 0 & 61.2 & 
100 & 0 & 100 & 
1 & 0 & 0.75 
\\
math.MathUtils &
1 &
1 & 
1 & 1 & 4 & 
100 & 0 & 100 & 
0 & 0 & 0 & 
0 & 0 & 0 
\\
lang.BooleanUtils &
2 &
5 & 
2 & 2 & 12 & 
50 & 100 & 83.3 & 
0 & 0 & 100 & 
0 & 0 & 0.9 
\\
guava.IntMath &
1 &
1 & 
3 & 3 & 46 & 
100 & 66.6 & 97.8 & 
100 & 0 & 0 & 
1 & 0 & 0 
\\
tsuite.Angle &
1 &
2 & 
3 & 1 & 0 & 
100 & 100 & 100 & 
0 & 0 & 0 & 
0 & 0 & 0 
\\
tsuite.MathUtil &
1 &
3 & 
3 & 1 & 13 & 
100 & 100 & 84.6 & 
33.3 & 0 & 33.3 & 
0.49 & 0 & 0.47 
\\
tsuite.Envelope &
1 &
1 & 
3 & 4 & 27 & 
100 & 100 & 18.5 & 
0 & 0 & 0 & 
0 & 0 & 0 
\\
\hdashline
eiffel.Composite &
1 &
4 & 
0 & 7 & 42 & 
100 & 100 & 50 & 
0 & 50 & 0 & 
0 & 0.66 &  0 
\\
eiffel.DLLN &
2 &
3 & 
0 & 4 & 29 & 
100 & 100 & 89.6 & 
0 & 66.6 & 100 & 
0 & 0.79 & 0.94 
\\
eiffel.Map &
3 &
6 & 
4 & 10 & 22 & 
50 & 100 & 81.8 & 
16.6 & 66.6 & 16.6 & 
0.24 & 0.79 & 0.27 
\\
eiffel.RingBuffer &
5 &
5 & 
9 & 31 & 269 & 
88.8 & 87 & 64.6 & 
20 & 40 & 20 & 
0.32 & 0.54 & 0.3 
\\
cozy.Polyupdate &
2 &
3 & 
3 & 3 & 119 & 
66.6 & 100 & 1.6 & 
33.3 & 66.6 & 66.6 & 
0.44 & 0.79 & 0.03 
\\
cozy.Structure &
2 &
2 & 
2 & 2 & 21 & 
100 & 100 & 95.2 & 
100 & 100 & 100 & 
1 & 1 & 0.97 
\\
cozy.ListComp02 &
2 &
2 & 
0 & 4 & 5 & 
100 & 100 & 83.3 & 
0 & 100 & 0 & 
0 & 1 & 0 
\\
cozy.MinFinder &
1 &
1 & 
0 & 2 & 2 & 
100 & 100 & 100 & 
0 & 100 & 100 & 
0 & 1 & 1 
\\
cozy.MaxBag &
3 &
3 & 
8 & 33 & 78 & 
100 & 84.8 & 94.8 & 
0 & 66.6 & 100 & 
0 & 0.74 & 0.97 
\\
\midrule
Total-65 & 
43 &
65 & 
65 & 131 & 975 & 
92.3 & 88.5 & 65.8 & 
27.6 & 38.4 & 52.3 & 
0.42 & 0.53 & 0.57 
\\
Total-SpecFuzzer & 
30 &
41 & 
43 & 76 & 618 & 
95.3 & 88 & 62.4 & 
41.4 & 43.9 & 82.9 & 
0.57 & 0.58 & 0.71 
\\
Total-GAssert & 
23 &
28 & 
39 & 54 & 615 & 
94.8 & 85.1 & 63.4 & 
64.2 & 46.4 & 71.4 & 
0.76 & 0.60 & 0.67 
\\
Total-EvoSpex & 
24 &
29 & 
34 & 82 & 624 & 
91.1 & 92.8 & 59.9 & 
44.8 & 86.2 & 72.4 & 
0.60 & 0.89 & 0.65 
\\
\bottomrule
\end{tabular}
\end{table*}

\subsubsection*{Inferred Assertions}

If we focus on the number of inferred assertions, GAssert and EvoSpex report fewer assertions than \tname{}. This is an advantage of the previous techniques, since the produced output is easier to interpret. The main reason here is that both techniques feature evolutionary processes, that aim at minimizing the size of the assertions (this is an objective of the evolution processes). \tname{} is in this respect a simpler technique. Still, the invariant selector component allows our tool to report a reasonable number of assertions (22 per method, on average). 

\subsubsection*{Precision}

Precision is the aspect in which both GAssert and EvoSpex outperform \tname{}. Again, this has to do with the fact that both GAssert and EvoSpex incorporate mechanisms to actively reduce the number of false positives (understood as invalid properties). In particular, GAssert iteratively improves assertions using OASIs~\cite{DBLP:conf/issta/JahangirovaCHT16}, launching EvoSuite instances to search and detect defects in the candidate assertions. EvoSpex uses bounded-exhaustive test generation technique with the aim of building a more thorough test suite, able to discard more false positives. Both techniques have disadvantages associated with these processes. GAssert pays a price in efficiency (it is the most costly of the three); EvoSpex's bounded exhaustive test generation has scalability issues (due to its bounded exhaustive test generation, it has difficulties scaling to larger subjects). 

\tname{} borrows the mechanism to deal with precision, that Daikon employs. This issue can be dealt with by improving test suite quality. We used Randoop in our experiments, which may be complemented by additional automated test generation techniques. 

\subsubsection*{Recall}

Recall is the aspect where \tname{} outperforms GAssert and EvoSpex. This is the case for the overall ground truth, and for most tool-specific ground truth subsets. EvoSpex has better recall than \tname{} for its specific language; it infers 5 assertions that \tname{} cannot. Out of these 5, 2 assertions are discarded by the invariant selector (we described the reasons above). The remaining three are supported by the grammar, but Daikon is currently unable to instantiate these assertions (we also described this issue earlier in the paper); that is, these three assertions are not part of the 41 that our prototype currently supports. 

The recall improvement of \tname{} over the other techniques makes our tool more effective overall, as summarized by the f1-scores. Notice that \tname{} has a better f1-score compared with previous techniques, for the overall ground truth, i.e., even taking into account its precision limitations, and the current issues with support for assertions.

\subsection{Threats to Validity}

Our experimental evaluation was performed on a data set built with subjects from previous works. We needed to manually study these subjects in order to define the ground truth assertions. To mitigate the risk of errors, we checked these assertions using Microsoft IntelliTest (previously named Pex~\cite{Tillmann+2008}). 

Threats to internal validity may arise from the randomness of the each technique. To account for this issue, we evaluated \tname{}, GAssert and EvoSpex over multiple runs on each subject method, and reported the averages. As further work, we plan to extend the experimental evaluation to larger-scale Java projects, which will likely imply abandoning the computation of performance metrics over ground truths, due to the effort that would involve studying larger projects and manually writing correct assertions. 

%% file: related-work.tex
\section{Related Work}
\label{relatedwork}



The use of assertions in programs has a long tradition. Originally, assertions were used as part of approaches for software verification \cite{DBLP:journals/annals/Hoare03}, and soon were incorporated into programming languages, for run-time checking ~\cite{DBLP:journals/sigsoft/ClarkeR06}. Assertions are currently used for multiple software development activities: program verification~\cite{DBLP:conf/fmco/ChalinKLP05,DBLP:conf/sas/Fahndrich10}, software design \cite{Meyer1997}, bug finding~\cite{Tillmann+2008,DBLP:conf/sigsoft/LeitnerCOMF07}, program comprehension and maintenance~\cite{DBLP:conf/icsm/SatpathySR04}, among others. 

Specification inference is an active area of research. Besides the techniques that infer contract assertions, with which we have compared our technique in this paper \cite{gassert2020,Molina+2021}, other related approaches exist, in particular for inferring test oracles \cite{DBLP:conf/issta/FraserZ10,DBLP:conf/icse/WatsonTMBP20} (that is, assertions that are valid only for specific unit tests), and other kinds of specifications, such as behavioral descriptions \cite{Caso2013,Lo2018,Lo2021}. These techniques seek related but different objectives, and thus can complement each other. In relation to the test oracle problem in particular~\cite{Barr+2015}, tools and techniques for inferring test assertions produce specifications that are difficult to generalize as contracts; contract specifications, on the other hand, can be instantiated as test assertions, but may capture weaker properties, compared to their test assertion counterparts. Other related techniques attempt to produce assertions from other sources, such as comments \cite{Blassi+2018}, or weaker forms of specifications, notably metamorphic relations \cite{BlasiGEPC2021}.

Our technique is strongly related to Daikon \cite{DBLP:journals/scp/ErnstPGMPTX07}, a mature tool/technique for likely invariant discovery. Our approach is largely motivated by automatically equipping Daikon with more expressive assertions. We are not aware of other approaches that automatically address the expressiveness limitations of Daikon. Fuzzing \cite{fuzzingbook2019:index} is also a very active topic, with known applications in security vulnerability discovery, and bug finding in general. To the best of our knowledge, our approach is the first to employ fuzzing to produce candidate formal specifications.

%% file: conclusion.tex
\section{Conclusion and Future Work}

Formal class specifications have applications in various areas of software development, including software design, bug finding, and program comprehension. Technique to automatically infer class specifications have been proposed, but are limited, e.g., they support a limited number of assertion types and are inflexible to change.


We present \tname{}, a technique to infer likely class specifications by combining static analysis, grammar-based fuzzing, and mutation analysis. Our evaluation shows that \tname{} has superior performance in comparison with the state-of-the-art tools GAssert and EvoSpex, especially considering recall. Furthermore, the use of grammar-based fuzzing enables \tname\ to be easily adapted to different assertion languages. 


This paper also opens various lines for improvement, as we have identified some concrete limitations of our approach. The mutation-based mechanism to discard weak assertions is affected by the absence of mutation operators, that would allow our tool to detect some specific constraints. Other more sophisticated mechanisms to deal with assertions not killed by any mutant may also be incorporated (e.g., constraint-based techniques). In general, the modular structure of our technique enables us to improve specific components, e.g., test generation (to improve precision), fuzzing (to consider more effective/efficient fuzzing techniques), etc. Finally, implementation limitations in the dynamic detection phase constitute a bottleneck for \tname{}'s assertion inference capabilities, that we plan to address in future extensions of our tool. The evaluation artifacts of \tname\ and a detailed description of how to reproduce our experiments are publicly available at the following link: \repoURL{}.

\newpage